\documentstyle[epsf]{l-aa}
\newcommand{\eqb}{\begin{eqnarray}}
\newcommand{\eqe}{\end{eqnarray}}
\newcommand{\kpar}{\kappa_{\scriptscriptstyle \|}}
\newcommand{\dmu}{\langle(\Delta\mu)^2\rangle}
\newcommand{\md}{\mbox{d}}
\newcommand{\dom}{\Delta\Omega_{\scriptscriptstyle \rm max}}
\newcommand{\dmumin}{\delta\mu_{0}}
\newcommand{\cdom}{\cos(\Delta\Omega_{\scriptscriptstyle \rm max})}
\newcommand{\mucr}{\mu_{\scriptscriptstyle \rm crit}} 
\newcommand{\alcr}{\alpha_{\scriptscriptstyle \rm crit}} 
\newcommand{\phidht}{\Phi_{\scriptscriptstyle \rm dHT}}
\newcommand{\ncross}{N_{\scriptscriptstyle \rm cross}}

\begin{document}
\thesaurus{12(02.01.1; 02.19.1; 03.13.4; 09.03.2; 09.19.2)}
\title{Particle acceleration at oblique shocks and discontinuities of the
 density profile}
\author{U. D. J. Gieseler\inst{1}\thanks{{\it Present address:\/} 
        University of Minnesota, Department of Astronomy,
        116 Church St. S.E., Minneapolis, MN 55455, U.S.A.}
 \and J. G. Kirk\inst{1} 
 \and Y. A. Gallant\inst{2} \and A. Achterberg\inst{2}}
\institute{Max-Planck-Institut f\"ur Kernphysik,
Postfach 10 39 80, 69029 Heidelberg, Germany
\and Sterrenkundig Instituut Utrecht, Postbus 80 000, 3508 TA Utrecht, 
Netherlands}
\offprints{gieseler@msi.umn.edu}
\date{Received \dots}
\maketitle
\begin{abstract}
In the theory of diffusive acceleration at oblique shock fronts the question 
of the existence of a discontinuity of energetic particle density is
contentious. The resolution of this problem is interesting from
a theoretical point of view, and potentially for
the interpretation of observations of particle densities at heliospheric 
shocks and of 
high-resolution radio observations of the rims of
supernova remnants. It can be shown analytically that an isotropic particle
distribution at a shock front implies continuity of the particle
density -- whether or not the shock is oblique. However, if the
obliquity of the shock induces an anisotropy, a jump is
permitted. Both semi-analytic computations and Monte-Carlo simulations
are used to show that, for interesting parameter ranges, a jump is
indeed produced, with accelerated particles concentrated in a
precursor ahead of the shock front.
\end{abstract}
\section{Controversy -- general description}
\label{contro}
In the test-particle theory of diffusive shock acceleration, the phase-space 
spectral index $s$ of accelerated particles depends solely on the 
compression ratio $r=\rho'/\rho$ of the shock (where $\rho$ and $\rho'$ 
are the upstream and downstream densities respectively):
$s=3r/(r-1)$, which results in $s=4$ for a strong shock in an ideal 
gas with $c_{\scriptscriptstyle p}/c_{\scriptscriptstyle V}=5/3$
(Axford et al.~\cite{AxLeSk77}; Krymskii~\cite{Krym77}; 
Bell \cite{Bell78}; Blandford \& Ostriker~\cite{BlOs78}). This result, like 
many other
analytical predictions (for a review see Drury~\cite{Drur83}; 
Blandford \& Eichler~\cite{BlEi87}), depends on the assumption that 
the phase space density is close to being isotropic, even at the  
shock front. In this case, the density profile of accelerated test particles 
is a continuous function of position 
(e.g., Kirk et al.~\cite{KiMePr94}, page~262). In planar symmetry, the density
is constant downstream and drops off exponentially upstream of the shock.
This situation applies to both parallel and oblique shocks, as was pointed 
out by, for example, Axford~(\cite{Axfo81}).

Recently, a discussion has arisen in the literature concerning the occurrence
of discontinuities in the density of accelerated particles at an oblique 
shock front. Whereas Ostrowski~(\cite{Ostr91}) (hereafter O91) finds a 
substantial effect, Naito \& Takahara (\cite{NaTa95}) (hereafter N\&T95) 
assert that the density is continuous.
Both of these papers present Monte-Carlo simulations of particle acceleration
in which the velocity of the shock front is a significant fraction of the 
speed of light, and take explicit account of a possible anisotropy of the 
particle distribution. From the theory of diffusion, it is well-known that 
the anisotropy of particles of speed $v$ is of the order of $u/v$, where $u$ 
is the speed at which the shock sweeps through the medium responsible for 
making the particles diffuse. At an oblique shock front, the speed
relevant for particles diffusing along a magnetic field line is the
velocity of the intersection point of the shock front and a given field line, 
$u=u_{\rm s}/\cos\Phi$, where $u_{\rm s}$ is the inflow speed along the
shock normal and $\Phi$ is the angle between the shock normal and the magnetic
field, measured in the upstream rest frame of the plasma. 
Thus, if the field is oblique, even a relatively slow shock front can produce 
a substantial anisotropy. The question which we address in this paper is 
whether or not this anisotropy leads to a discontinuity in the particle 
density and under what conditions such an effect could be observed.

An approximation which is often used in treating oblique shocks is that in
which a particle crossing the shock conserves its magnetic
moment (also referred to as the first adiabatic invariant; 
e.g., Webb et al.~\cite{WeAxTe83}).
This approximation is valid for non-relativistic
perpendicular shocks (Whipple et al. \cite{whippleetal86}). 
Terasawa (\cite{Tera79}) and 
Decker (\cite{Deck88}) have performed numerical simulations which confirm its 
accuracy at non-relativistic oblique shocks and
Begelman \& Kirk (\cite{BeKi90}) have conducted tests
at relativistic shocks. It is important to bear in mind, however, that this 
approximation breaks down for 
sufficiently fast shocks (see, for example, 
Kirk et al.~\cite{KiMePr94}, page~241). 
Conservation of magnetic moment implies
that a particle can be reflected by the magnetic compression at a
fast-mode shock front, and the question of the existence of a discontinuity 
in the particle density is intimately connected with the
phenomenon of reflection, as pointed out in an early paper
on this subject (Achterberg \& Norman \cite{AcNo80}). 
If this approximation is adopted, the problem of acceleration at an oblique 
shock of particles which undergo pitch-angle diffusion along field lines can 
be solved semi-analytically (Kirk \& Heavens \cite{KiHe89}, 
hereafter K\&H89), at least for the case in which the accelerated
particles are ultra-relativistic ($v=c$) and build a power-law distribution 
in momentum. In Sect.~\ref{monte}, we address the question of the existence of
a discontinuity using a new Monte-Carlo code incorporating the assumption of
conservation of magnetic moment (Gieseler \cite{Doc98}). 
Since there already exist contradictory simulations in the literature, we take
particular care to check the results obtained with this new code against the 
analytic results of K\&H89. Both the code and the analysis find a 
discontinuity. However, although N\&T95 
also adopted the assumption of conservation of magnetic moment
in their simulations,
they did not find a discontinuity. We show that this can be caused by low
spatial resolution of the simulation and does not necessarily invalidate
results concerning the spectrum of accelerated particles. 
Ostrowski~(\cite{Ostr91}), 
on the other hand, {\em did} find a discontinuity, without implementing the
conservation of magnetic moment. In Sect.~\ref{analytical} we argue that 
Liouville's theorem in fact guarantees continuity in this case.
This result can once again be caused by a rapidly varying density profile
which is sampled at low spatial resolution; in this case it appears because 
the length scale of a gyration radius 
was not resolved. However, Ostrowski's results do indicate that the 
accumulation of
particles upstream of a relativistic shock front is an important physical
effect, even though there may not be a discontinuity in the strict sense.

At non-relativistic shocks, such as those observed in the solar system, the
analytic method of K\&H89 fails, and one must rely on simulations. 
Here again the situation is not clear-cut.
Monte-Carlo simulations by Ellison et al.~(\cite{ElBaJo96}) do not show 
discontinuities or a jump across the shock front, whereas recent 
numerical solutions using the finite difference method have found 
such effects (Ruffolo~\cite{Ruff99}).
In Sect.~\ref{dens_profile} we
present high resolution test-particle simulations using parameters 
appropriate for solar system shocks and demonstrate the existence of 
accumulated particles in the precursor of the shock.

The resolution of this question is not only of formal interest, but is also
relevant for the interpretation of data taken by the Ulysses spacecraft and
of observations of the radio emission of supernova remnants. These applications
are discussed briefly in Sect.~\ref{disc}.

\section{Analytical considerations}
\label{analytical}
The system we consider consists of energetic particles which move in two
half-spaces separated by a shock front. 
Plasma flows into and out of the shock front carrying with it a uniform 
magnetic field. In addition to the uniform 
field, we assume there exist magnetic fluctuations static in the rest frame 
of the plasma on each side of the shock whose 
effect on the particle motion may be described via a scattering operator. Two
levels of approximation are 
important. In the first, the particle is described by its full six-dimensional
phase space coordinates. The trajectory 
is integrated in an explicit realisation of the stochastic magnetic field,
taking full account of the gyration phase. 
Upon crossing the shock front, the trajectory experiences no forces other than
those exerted by the magnetic field. 
This is the approach used by O91. The second level of 
approximation is one in which the particle is described by five coordinates: 
the position of the guiding centre, the 
magnitude of the momentum and the pitch angle. The magnetic field is assumed
uniform and the guiding centre follows a field line except when 
crossing the shock. The magnetic moment is conserved in between scattering 
events, which are assumed to change 
only the pitch angle, and also upon crossing the shock front. This is the
approximation used by K\&H89 and N\&T95. It is generally referred to in the 
literature as the \lq drift 
approximation\rq, and we adopt this terminology here, although we do not 
explicitly consider the drifts themselves.

In both levels of approximation a discontinuity can in principle arise in the 
formal 
description of the dependence of the distribution function on the spatial
coordinate along the shock normal. However, the physical interpretation of a 
mathematical discontinuity is different in each case. 

In the first approximation, the shock front is taken to have zero thickness. 
In reality, the shock transition will extend 
over a finite region in space, which is assumed to be small compared to the
other length scales of interest, for example the gyration radius of the 
energetic
particles. Thus, if the acceleration process were to produce a significant 
change in the energetic particle density over a
length scale smaller than the gyration radius, but perhaps comparable to the 
shock
thickness, this would appear as a discontinuity in simulations such as
those of O91. 
However, since particles do not suffer impulsive deflection at the shock 
front, i.e., the momentum coordinates of a particle are in general
continuous functions of position even at the shock front itself, 
it follows from Liouville's theorem that the phase-space distribution
function, which is constant along trajectories, is also a
continuous function of position across the shock front. The particle
density is simply an integral of this function over all momenta, so that it is
also continuous, provided the momentum is measured in the same frame of
reference both upstream and downstream of the shock. Thus, there can be no
formal discontinuity in this approach. Of course, the 
density may vary smoothly on the length scale of the gyration radius, so that 
we can interpret Ostrowski's result as due to a strong 
gradient in the density on this length scale.

The drift approximation, however, can only resolve changes which occur on
length scales longer than a gyration radius, so that the distribution found by 
Ostrowski {\em must} appear as a discontinuity in simulations which use this 
approximation. There is a close relationship between such a discontinuity 
and the angular distribution, which 
can be understood as follows. Consider an
oblique shock viewed in the de~Hoffmann-Teller (hereafter dHT) frame (de~Hoffmann \&
Teller~\cite{deHoTe50}; Kirk et al.~\cite{KiMePr94}), in which the electric
field vanishes and the shock is stationary. A particle trajectory is now
described by only five coordinates, of which the magnetic moment is conserved
both between scatterings and on encountering the shock front.
Denoting by $p_{\scriptscriptstyle \bot}$ the component of the
particle momentum perpendicular to the magnetic field $\vec{B}$, the magnetic moment is 
$p_{\scriptscriptstyle \bot}^2/B$.
Because $p=|\vec{p}|$ is also conserved in the dHT frame 
(in which the electric field vanishes) this leads to 
\eqb\label{adiabat}
\frac{1-\mu^2}{B} = \frac{1-(\mu')^2}{B'}\,,
\eqe
where a prime denotes downstream quantities. Rearranging, the downstream 
pitch angle is given by
\eqb
\mu'&=&
\left(\mu/|\mu|\right)\sqrt{(\mu^2-\mucr^2)/(1-\mucr^2)}\,,
\eqe
where the cosine of the of the loss-cone angle is given by
 $\mucr=\sqrt{1-B/B'}$. Thus, according to its pitch angle, an upstream 
particle may be reflected or transmitted, and we can divide phase space
on the upstream side of the shock into four regions:
\begin{enumerate}
\item
 $\mucr<\mu<1$, particles approaching the shock which will be transmitted 
(i.e., in the loss cone);
\item
 $0<\mu<\mucr$, particles approaching the shock which will be reflected;
\item
 $-\mucr<\mu<0$, particles leaving the shock after reflection;
\item
 $-1<\mu<-\mucr$, particles leaving the shock after transmission
from downstream.
\end{enumerate}
The phase space downstream splits into just two regions,
$\mu'>0$ and $\mu'<0$, since no trajectories incident from downstream are 
reflected.

In the drift approximation, the phase-space distribution function is
independent of gyration phase to lowest order (e.g., Spatschek~\cite{Spat90}, 
page 145), so that application of Liouville's theorem yields
\eqb
f(p,\mu)&=&f(p,-\mu)\quad {\rm for\ }|\mu|<\mucr \quad {\rm (reflection)}\,,
   \label{liou_refl}\\
f(p,\mu)&=&f'(p,\mu')\quad {\rm for\ }|\mu|>\mucr \quad {\rm (transmission)}\,,
   \label{liou_trans}
\eqe
where conservation of the momentum $p$ in the dHT frame is used.

The upstream and downstream densities are given by
\eqb
n &=& \int\limits_{-1}^{-\mucr}\!\!\md\mu\, n_{\rm t}(\mu)
  +\int\limits_{-\mucr}^{\mucr}\!\!\md\mu\, n_{\rm r}(\mu)
  +\int\limits_{\mucr}^{1}\!\!\md\mu\, n_{\rm t}(\mu) \,,\label{n_int}\\
n'&=&  \int\limits_{-1}^{1}\!\md\mu'\, n'_{\rm t}(\mu')\,,\label{n_int_2}
\eqe
where we have defined the quantities 
\eqb
n_{\rm t,r}(\mu)&:=&\int f(p,\mu)\,2\pi\,p^2\,\md p\,,\label{n_tr}\\
n'_{\rm t}(\mu')&:=&\int f'(p,\mu')\,2\pi\,p^2\,\md p\,= n_{\rm t}(\mu)\,, 
                                                  \label{n_prime_t}
\eqe
such that the suffix r refers to particles which are or will be
reflected (i.e., $|\mu|<\mucr$) and the suffix t to ones which are or
will be transmitted ($|\mu|>\mucr$, or $-1<\mu'<1$). 
The second relation in Eq.~(\ref{n_prime_t}) follows from 
Eq.~(\ref{liou_trans}).

A continuous density distribution at the shock front ($n=n'$)
is in general not guaranteed. 
This can be seen by using specific assumptions
about the density of transmitted and reflected particles and 
Eqs.~(\ref{n_int})--(\ref{n_int_2}) (Gieseler~\cite{Doc98}).
For the physical distribution of Fig.~\ref{pitch}, the values of the density
which can be calculated from Eq.~(\ref{n_int}) and Eq.~(\ref{n_int_2})
are quite different, as discussed below. The reason for the upstream and 
downstream densities to be different can be understood from 
Eq.~(\ref{n_prime_t}), which preserves a \lq balance\rq\ of the transmitted 
particles, whereas the reflected particles contribute only to the 
upstream density. 
A continuous density is obtained if there is no compression of the magnetic 
field ($B=B'$). In this case no particles are reflected ($\mucr=0$) and 
Eqs.~(\ref{n_int})--(\ref{n_prime_t}) then give $n=n'$. 
This is valid for a parallel shock and the trivial case in which no shock 
front is present.
If the pitch-angle distribution is isotropic at an oblique shock front, 
then one again finds $n=n'$ by integration of Eqs.~(\ref{n_int}) and 
(\ref{n_int_2}). Noting that in this case $n_{\rm t}(\mu)$ and
$n_{\rm r}(\mu)$ have the same constant value,
continuity follows from Eq.~(\ref{n_prime_t}). In other words, the contribution
of the reflected particles upstream exactly balances the compression of the
transmitted ones downstream in this case.

To summarise this section, in general the density of accelerated test particles
at oblique shocks will vary on the length scale of the gyration radius across 
the shock front. This variation is closely connected with the anisotropy of the
angular distribution. It should appear as a discontinuity in treatments which
use the drift approximation to the particle motion. 
However, when conditions are such that the theory of diffusive
acceleration applies, i.e., the particle velocity is 
much larger than the shock speed projected
along the magnetic field ($v\gg u_{\rm s}/\cos\Phi$), then the anisotropy and
the associated density variation are small.
\section{Monte-Carlo simulations}
\label{monte}
We now present results from test-particle simulations of accelerated 
particles at shock fronts. 
The key aspects of the technique have been used and described by several 
authors (e.g.,  Kirk \& Schneider~\cite{KiSc87}; O91; 
Baring et al.~\cite{BaElJo93}; N\&T95),  
so that a brief description suffices.
%
%
We consider oblique shocks, where the magnetic field is inclined at
an angle $\Phi$ with respect to the shock normal in the upstream rest frame, 
and has no dynamical effect on the plasma flow. 
The shock speed in this frame is $u_{\rm s}$. We consider (in principle) the 
whole range of $\Phi$ for subluminal shocks, e.g.
 $u_{\rm s}\le u_{\rm s}/\cos\Phi < 1$ (here and below: $c=1$).
The gyration centre of a particle's trajectory is followed in the upstream 
and downstream rest frames of the background plasma.
%
%
In these frames, as in the dHT frame, the momentum $p=|\vec{p}|$ is constant. 
Particles move 
along the magnetic field under the influence of small scale 
irregularities which lead to pitch-angle scattering.
We do not investigate the effect of transport of particles perpendicular 
to the mean magnetic field.  We use an algorithm
for calculating a pitch angle $\mu_{\scriptscriptstyle \rm new}$ from a 
given pitch angle $\mu$ which was given by O91 (see Fig. 1 therein).
From two random numbers $R_1$ and $R_2$ which are uniformly distributed on 
the interval $[0,1]$, the new pitch angle is given by
\eqb
\cos\Delta\Omega &=& 1-(1-\cos\Delta\Omega_{\scriptscriptstyle \rm max})
                         \,R_1\,,\label{scatt_1}\\
\mu_{\scriptscriptstyle \rm new} &=& \mu\,\cos\Delta\Omega + \sqrt{1-\mu^2}
    \sin\Delta\Omega\,\cos(2\pi R_2)\,,\label{scatt_2}
\eqe 
where we have chosen $\dom=0.1$ for most of the simulations shown below.
This gives a very good approximation of pitch-angle scattering with an 
infinitesimal amplitude. The results (at least for the spectral index and 
the density distribution) do not change substantially for a factor of 5
higher or lower value of $\dom$.\footnote{O91 used $\dom=0.3$ as an 
approximation for infinitesimal pitch-angle scattering.}
The time step $\Delta t$ for successive scatterings is kept constant.
These scatterings are performed in the upstream and downstream rest frames.
%
%
At a shock crossing, we transform the particle momentum $p$ and the pitch angle
 $\mu$ into the dHT frame, which is always possible at subluminal oblique 
shocks (for a description of the Lorentz transformations between these frames 
see K\&H89 or Gieseler~\cite{Doc98}), and apply the conservation of magnetic 
moment (Eq.~\ref{adiabat}) to find the new downstream pitch angle in the 
dHT frame. Transformation into the downstream rest frame then gives 
the new values of the momentum and pitch angle. This method is also 
used by K\&H89 and N\&T95, to which we compare our results. 

The validity of the assumption of conservation of the magnetic moment can be
evaluated as follows. First, consider a \lq strict\rq\ condition which
can be applied to a single particle crossing the shock 
(Kirk et al.~\cite{KiMePr94}, page~241). The number of times a trajectory
intersects the shock front can be estimated as 
\eqb\label{strict}
\ncross \approx \tan\alpha\,\tan\phidht\,,
\eqe
where $\alpha=\cos^{-1}\mu$ is the pitch angle, and
 $\phidht$ is the angle between the upstream field and the shock normal 
measured in the dHT frame.\footnote{Note that
 $\cos^2\Phi=(1+u^2_{\rm s}\tan^2\phidht)/(1+\tan^2\phidht)\,$.\label{dht_trans} }
Adiabatic behaviour is expected when \mbox{$\ncross\gg 1$}.
An accurate separation between transmitted and reflected particles
is guaranteed if we demand this condition to be valid for at 
least all particles outside the loss cone, i.e.,
 $|\mu|\le\mucr\equiv\cos\alcr =\sqrt{1-B/B'}$, where
 $B/B'=[(1+\tan^2\phidht)/(1+r^2\tan^2\phidht)]^{-1/2}$ and $r$ is the
compression ratio. Inserting the
critical pitch angle $\alcr$ in Eq.~(\ref{strict}) and expanding in powers of
 $1/\ncross\,$, we get a condition
for the upstream inclination angle in the dHT frame which is independent
of the fluid speed:
\eqb\label{condition}
\tan\phidht\gg\,\sqrt{r-1}\,.
\eqe
For the non-relativistic shock speed $u_{\rm s}=0.01$, Eq.~(\ref{condition}) 
is fulfilled for
essentially the whole parameter range in $u_{\rm s}/\cos\Phi$ considered here.
For shock speeds in the mildly relativistic region, Eq.~(\ref{condition}) 
always holds in an (albeit small) region at \mbox{$u_{\rm s}/\cos\Phi \to 1$}.

The condition $\ncross\gg 1$ is, however, much too strict. A less conservative
approach is to assume validity of the approximation in an average (over
phases) sense. This has been shown by simulations as mentioned above.
For example Decker~(\cite{Deck88}) finds that for
non-relativistic shocks the results compare well with adiabatic
behaviour over the whole range of $\Phi$. In particular, the data shown in
his Fig. 8 show close agreement concerning the reflection probability for
the approximate and numerical results.
Ostrowski~(\cite{Ostr91}) used an extended Monte-Carlo code which does not
assume adiabatic behaviour. He was thus able to check the quality of the
approximation at mildly relativistic shock speeds.
His Fig. 3 shows good agreement with K\&H89 for the spectral index. 
Departures start to occur at $u_{\rm s}=0.5$, but always diminish as
$u_{\rm s}/\cos\Phi\rightarrow1$. Furthermore, we have directly compared our
simulations with his for $u_{\rm s}=0.3$ at $\Phi=60^{\circ}$ and $70^{\circ}$
and found the same density discontinuity (see below).
\begin{figure}[t]
    \vspace{-3.3cm}
    \begin{center} 
      \epsfxsize9.0cm 
      \mbox{\epsffile{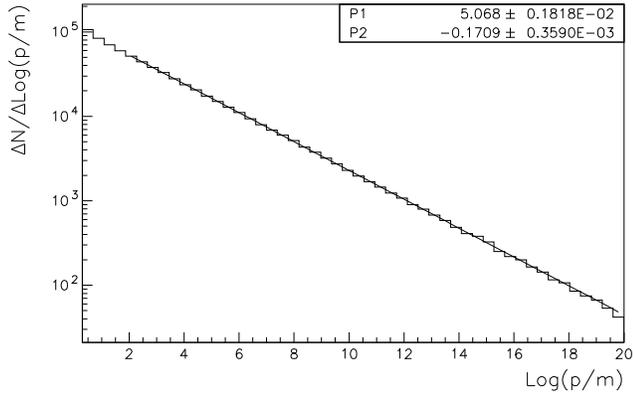}}
    \end{center}
    \vspace{-1.4cm} 
    \protect\caption{Monte-Carlo simulation of a momentum spectrum for 
             compression ratio $r=4$, shock speed $u_{\rm s}=0.4$ and an 
             angle of the magnetic 
             field of $\Phi=45^{\circ}$. The speed of the intersection of
             magnetic field and shock is then $u_{\rm s}/\cos\Phi=0.5657$
             (same parameters as for Fig.~\protect\ref{pitch} and 
             Fig.~\protect\ref{profile}). The spectral index of the phase
             space density is $s=3.1709\pm 0.0004$.}
    \label{impuls}
\end{figure}

%
%
To test the accuracy of our scheme against the calculation of K\&H89, 
particles are
injected in the upstream plasma with a velocity corresponding to
 $\gamma v = p/m \ge 2$, which means that they are already relativistic at 
injection. For intersection speeds of shock and magnetic field 
of $u_{\rm s}/\cos(\Phi)\ge 0.9$ the injection 
momentum has to be even higher. We are interested in particles
with momenta where the distribution has attained a pure power law, and
especially in the spectral index of this power law. 
Only particles from the power-law region in the 
momentum distribution are used for the pitch-angle and density distributions. 
Within this region we found that the angular distribution is independent of 
the momentum range chosen, and the power-law index is thus independent of
angle.
Simulation of an individual particle terminates when a maximum momentum
is reached (two orders of magnitude higher than the upper limit of 
Fig.~\ref{impuls}), or when the particle reaches a distance from the shock on
the downstream side which is three times greater than the left boundary of 
Fig.~\ref{profile}, at which the
density distribution has already reached its constant downstream value.
These limits vary with the parameters $u_{\rm s}$ and $\Phi$, and 
were always chosen such that boundary effects are not important.
The results from Monte-Carlo simulations and semi-analytical 
computations are shown (with the exception of Fig.~\ref{dens_non_rel})
for a compression ratio of $r=4$.

The three main aspects of the results are spectral index, angular distribution
and spatial distribution, and these are presented in turn.
\subsection{Spectral index}
\label{spectral_index}
\begin{figure}[t]
    \vspace{-3.1cm}
    \begin{center} 
      \epsfxsize9.0cm 
      \mbox{\epsffile{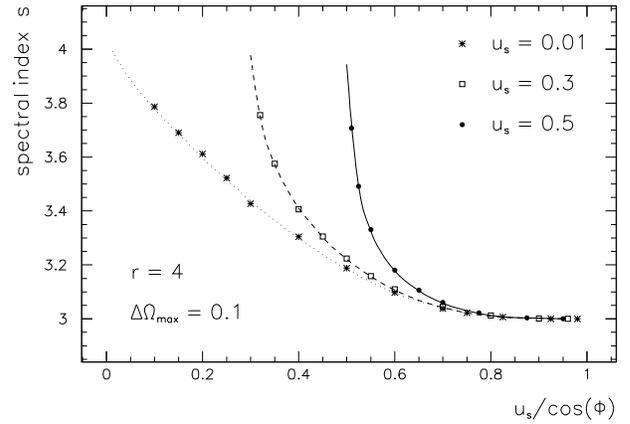}}
    \end{center}
    \vspace{-1.3cm} 
    \caption{Comparison of the spectral index from Monte-Carlo simulations
 for infinitesimal pitch-angle scattering with
 $\dom =0.1$
 (dots, stars, squares) with semi-analytical results from K\&H89 (solid line:
  $u_{\rm s}=0.5$; dashed line: $u_{\rm s}=0.3$; dotted line:
  $u_{\rm s}=0.01$).}
    \label{index}
\end{figure}

The momentum distribution of particles is measured 
at the shock front in the upstream rest frame. An example 
is shown in Fig.~\ref{impuls}, where $10^6$ independent particles
were simulated. The plot also contains a fit to 
the function $y=10^{p_1+p_2 x}$, together with the values of the
fit parameters and their statistical errors. 
The spectral index can be calculated from
 $\Delta N/\Delta \log(p/m)\propto p^{-s+3}$; this gives $s=3-p_2$.
The statistical error of the fit to the spectra is less than $1\%$ for
all spectral indices discussed below. 
To achieve maximum accuracy from the Monte-Carlo simulations, we fit the 
spectrum over a large finite range of momentum and did not include loss 
mechanisms.  
For $s<3.2$ the statistical errors are less than $0.1\%$.
For spectral indices near $s\simeq 3$ the statistics are very much better, 
because particles then gain energy mainly due to reflection. 
This means that not much time is 
taken in following the particle trajectories in those parts of 
the downstream region where the particle has a low chance of returning to the
shock front. As a result, the error in $s$ is less than 
 $0.01\%$ for spectral indices corresponding to $u_{\rm s}/\cos\Phi >0.8$.
Fig.~\ref{index} shows the spectral indices for non-relativistic and mildly
relativistic shock velocities and a wide range of inclination angles $\Phi$. 
The lines in this plot are taken from Fig.~2 and Fig.~6 of K\&H89 and are in 
precise agreement with the simulations. 

\begin{figure}[t]
    \vspace{-3.1cm}
    \begin{center} 
      \epsfxsize9.0cm 
      \mbox{\epsffile{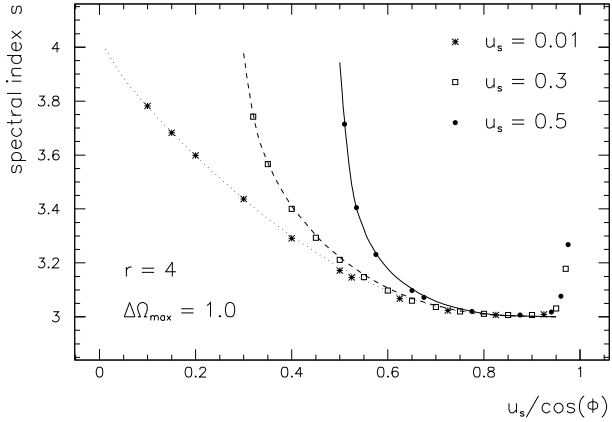}}
    \end{center}
    \vspace{-1.3cm} 
    \caption{Comparison of the spectral index for {\em intermediate} 
   pitch-angle scattering with
  $\dom =1.0$  from Monte-Carlo simulations (dots, stars, squares) with 
  {\em infinitesimal} pitch-angle scattering (lines) which are 
 semi-analytical results from K\&H89 (solid line: $u_{\rm s}=0.5$; 
 dashed line: $u_{\rm s}=0.3$; dotted line: $u_{\rm s}=0.01$).}
    \label{index_do1}
\end{figure}

Very important for an understanding of the spectral index and the density 
profile is the effect of 
the underlying scattering law. The results become quite different for
isotropisation after each scattering, given by $\dom = \pi$.
Although this does not correspond exactly to the simulation of 
large-angle scattering on point-like scattering centres 
(for which it would be necessary to choose exponentially distributed time 
steps, and consider cross field diffusion), the effects it produces should be
qualitatively similar. Using a scattering law in the transition regime
between infinitesimal pitch-angle scattering and isotropisation 
after each scattering allows an investigation of this dependence.
We focus on one example with 
 $\dom =1.0$ (intermediate pitch-angle scattering). This kind of scattering 
can have a very big influence on the 
precursor of accelerated particles. Under pitch-angle scattering with
 $\dom \ga 1$ particles have a higher escape probability from the 
shock because they are free to change their pitch angle to a value in the
loss cone ($\mucr<\mu<1$) within a few scatterings, and may then be 
transmitted through the shock.
This reduces the upstream density, as we will see later, but also
leads to a steeper spectrum for accelerated test particles. 
Figure~\ref{index_do1} shows a comparison of the spectral index
for intermediate and infinitesimal pitch-angle scattering. 
The stars, squares and dots show Monte-Carlo results for intermediate 
pitch-angle scattering with $\dom =1.0$, whereas the lines
are (as in Fig.~\ref{index}) results from K\&H89 for infinitesimal pitch-angle
scattering ($\dom \ll 1$). It can be seen that even for a relatively large 
value of $\dom =1.0$, the effect is very small for most intersection 
velocities of magnetic field and shock.

For infinitesimal pitch-angle scattering and $u_{\rm s}/\cos\Phi > 0.8$, 
a flat $s\simeq 3$ spectrum is achieved. 
In the standard picture, this spectrum corresponds to  
acceleration with vanishing escape probability. Here it is associated with
a very strong pile-up (see Fig.~\ref{jump_us}). A large ($\sim 1$) value
of $\dom$ reduces the pile-up (see Fig.~\ref{jump_us_do1}) 
and effectively increases the escape 
probability, leading to the steeper spectrum shown in Fig.~\ref{index_do1} 
for $u_{\rm s}/\cos\Phi>0.95$. 
%
%
For $\dom =\pi$ (isotropisation at each scattering) the spectrum becomes 
steeper for all inclination angles. This point was noted by N\&T95 
(their Fig.~7), and our results for the spectral index are in good agreement with 
theirs.
\subsection{Pitch-angle distribution} 
\label{pitch_dist}
\begin{figure}[t]
    \vspace{-2.9cm}
    \begin{center} 
      \epsfxsize9.0cm 
      \mbox{\epsffile{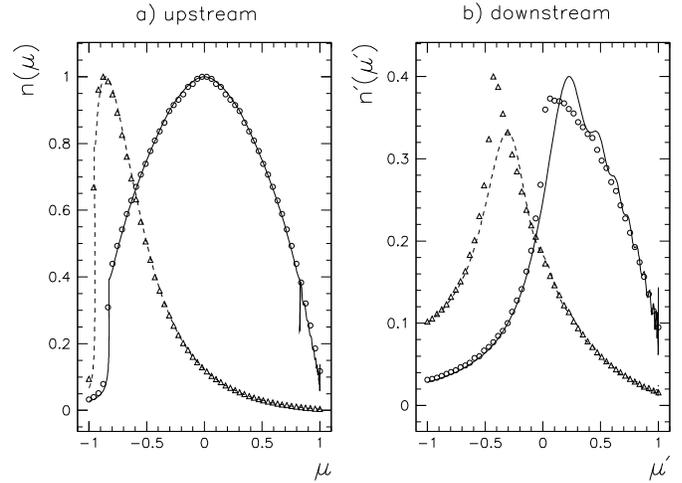}}
    \end{center}
    \vspace{-0.7cm} 
    \caption{Pitch-angle distribution at the shock front for
             $\dom=0.1$ and compression ratio $r=4$. 
             Figs. a) and b) show the distribution at the upstream and
             downstream side of the shock. In each figure the lines are
             from K\&H89 and the discrete symbols display the \lq bins\rq\ of 
             the Monte-Carlo simulations. The solid lines and the circles
             show the distribution in the dHT frame, and the dashed lines and
             the triangles show the distributions in the upstream (a) and 
             downstream (b) rest frame. See text for the normalisation. 
             The shock speed is $u_{\rm s}=0.4$, and the
             upstream inclination angle of the magnetic field
             $\Phi=45^{\circ}$ (compare Fig.~\protect\ref{impuls} and 
             Fig.~\protect\ref{profile}).}
    \label{pitch}
\end{figure}

As shown in Sect.~\ref{analytical}, the pitch-angle distribution of
accelerated particles plays a crucial role in determining whether
the density distribution has a jump at the shock front. For every 
set of parameters we have measured the pitch-angle distribution and present 
an example of the results in Fig.~\ref{pitch}.
The lines represent the distributions calculated semi-analytically by K\&H89,
whereas the discrete symbols show the contents of the \lq bins\rq\ filled in 
the above described Monte-Carlo method. The four representations of the 
pitch-angle distribution at the shock shown in Fig.~\ref{pitch} are:
1. Upstream of the shock in the upstream rest frame (Fig.~\ref{pitch}a dashed
line and open triangles). 2. Upstream of the shock in the dHT frame
(Fig.~\ref{pitch}a solid line and open circles). 3. Downstream of the shock in 
the downstream rest frame (Fig.~\ref{pitch}b dashed line and open triangles).
4. Downstream of the shock in the dHT frame (Fig.~\ref{pitch}b solid line 
and open circles). The upstream distributions (Fig.~\ref{pitch}a) are 
normalised to their maximum value. 
The normalisation at the downstream side of the shock are as follows:
the Monte-Carlo distribution in the dHT frame has the same normalisation
as the corresponding upstream one, to allow a direct comparison
between the two (see below); however, the 
Monte-Carlo distribution in the downstream rest frame 
is normalised to a maximal value of 0.4 to enable it to be displayed in the 
same figure. In each case, the normalisation of the semi-analytical 
results of K\&H89 are chosen to provide the best fit to the Monte-Carlo 
distributions. 
Comparison of the pitch-angle distributions from the two methods 
in Fig.~\ref{pitch} confirms that they are in close agreement. 

Particles with pitch angle $\mu>0$ in the dHT frame move in the downstream 
direction.
From Eq.~(\ref{adiabat}) we can calculate a critical pitch angle cosine 
below which these particles will be reflected upstream of the shock. 
For $r=4$, $\Phi=45^{\circ}$ and $u_{\rm s}=0.4$ we get $\mucr=0.826$.
Reflected particles contribute symmetrically about $\mu=0$ and constitute the 
major part of the distribution
(see solid line of Fig.~\ref{pitch}a), indicating that repeated reflections 
are effective in keeping particles upstream. The particles with
 $\mu<-\mucr=-0.826$ in the dHT frame are 
particles which cross the shock from the downstream side.
A Lorentz transformation\footnote{For particles with velocity $v=c$.} 
of the critical angle  $-\mucr=-0.826$ into the 
upstream rest frame gives $(-\mucr)^{\ast} = -0.949$.
At the downstream side of the shock the distribution in the dHT frame 
(solid line of Fig.~\ref{pitch}b) is simply divided into particles
going upstream ($-1\le\mu<0$) and those going downstream ($0<\mu\le 1$).
The boundary $\mu_{\scriptscriptstyle \rm b}=0$ between these regions is 
transformed to $\mu_{\scriptscriptstyle \rm b}^{\star}=-0.446$ in the 
downstream rest frame.
\subsection{Density profile}
\label{dens_profile}
From the pitch-angle distributions one can calculate the density using
Eqs.~(\ref{n_int}) and (\ref{n_int_2}), which is the integration of the 
pitch-angle distributions in the dHT frame. From Fig.~\ref{pitch} we see that 
the distribution immediately downstream of the shock
is $\la 0.38$ (open circles in Fig.~\ref{pitch}b), compared to $\le1$ for the
upstream distribution (open circles in Fig.~\ref{pitch}a). 
By comparing the upstream and downstream distributions in the
dHT frame, it is obvious that the downstream integral is less than upstream,
and therefore the density is discontinuous at the shock.
\begin{figure}[t]
    \vspace{-3.1cm}
    \begin{center} 
      \epsfxsize9.0cm 
      \mbox{\epsffile{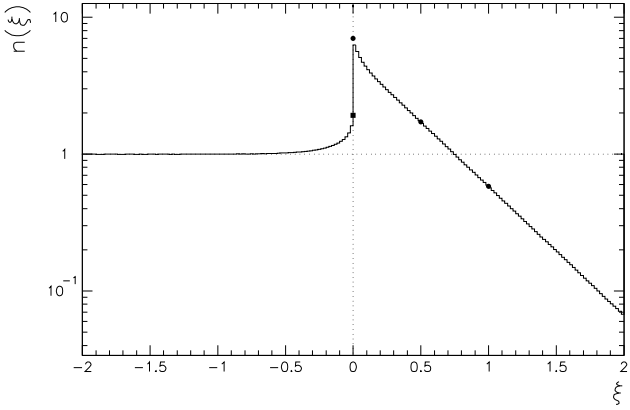}}
    \end{center}
    \vspace{-1.3cm} 
    \protect\caption{Monte-Carlo Simulation of the density profile at an 
            oblique shock with compression ratio $r=4$, maximal scattering 
            angle $\dom=0.1$, shock speed $u_{\rm s}=0.4$
            and inclination angle $\Phi=45^{\circ}$ (compare 
            Fig.~\protect\ref{impuls} and Fig.~\protect\ref{pitch}). 
            The solid line shows the \lq bins\rq\ of a position measurement of 
            particles, whereas the filled dots show a density measured 
            through the flux through a surface with constant distance upstream
            of the shock. The filled square indicates a measurement downstream
            of the shock.}
    \label{profile}
\end{figure}

In planar symmetry, the density distribution of accelerated test particles
is a function of the distance from 
the shock front. We use the 
dHT frame in order to compare the densities upstream and downstream 
directly, as in Sect.~\ref{analytical}. In this frame the shock is stationary
(at $x=0$). We normalise the distance $x$ perpendicular to the shock as the 
dimensionless variable $\xi$ according to
\eqb\label{norm}
\xi = \frac{u_{\rm s}}{\kpar\,\cos^2\Phi}\,x\,,
\eqe
where $\Phi$ is the upstream inclination angle of the magnetic field as 
defined above, and $\kpar$ the parallel diffusion coefficient 
(e.g. Jokipii~\cite{Joki71}; Skilling~\cite{Skil75}; Decker~\cite{Deck88}):
\eqb\label{kappa}
\kpar =\frac{v^2}{4}
       \int\limits_{-1}^{1}\md\mu\,\frac{1-\mu^2}{\nu_{\rm s}}\,,\quad\,\,
       \nu_{\rm s}=\frac{1}{1-\mu^2}\frac{\dmu}{\Delta t}\,\,,
\eqe
where $\nu_{\rm s}$ is the scattering frequency and we have used $v=c$ in the 
normalisation.
Note that $\kpar$, while convenient for the determination of a length scale, 
does not describe the transport of 
particles for the anisotropic distributions discussed here.
The mean squared variation in $\mu$ can easily be calculated from 
Eqs.~(\ref{scatt_1}) and (\ref{scatt_2}). Defining $\dmumin:=\cdom$, then
\eqb
\dmu = \frac{1}{2}\big(1-\dmumin\big)\big(1-\dmumin\mu^2\big)
           - \frac{1}{6}\big(1 -\dmumin\big)^2\,.
\eqe
\begin{figure}[t]
    \vspace{-3.1cm}
    \begin{center} 
      \epsfxsize9.0cm 
      \mbox{\epsffile{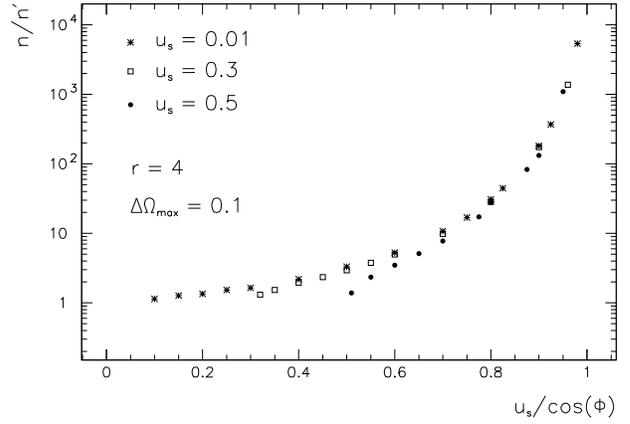}}
    \end{center}
    \vspace{-1.3cm} 
    \protect\caption{Ratio of the upstream to the downstream density at the 
             shock front vs. the intersection velocity of the shock and
             the magnetic field for {\em infinitesimal} pitch-angle scattering
             with $\dom =0.1$, compression ratio $r=4$ and three different 
             shock speeds. The values of $n/n'$
             are taken from flux measurements at the shock front,
             exemplified by the filled circle and square at $\xi = 0$
             in Fig.~\protect\ref{profile}.}
    \label{jump_us}
\end{figure}
Figure~\ref{profile} shows the steady-state density in the dHT frame for a 
shock
with velocity $u_{\rm s}=0.4$, compression ratio $r=4$ and inclination angle
 $\Phi=45^{\circ}$. The upstream plasma velocity in the dHT frame is 
then $u_{\rm s}/\cos\Phi=0.5657$. The plot shows the density 
 $n(\xi)$ as a function of the distance to the shock $\xi$. 
Contributions to this density were limited to particles from the 
power law part of the spectrum, indicated by the fit in Fig.~\ref{impuls}. 
Because the question of spatial resolution is of crucial importance to our
discussion, we use two independent methods to evaluate the density.
The solid line shows the contents of spatial \lq bins\rq, where 
particles simply contribute after every time step to the \lq bin\rq\ at their 
actual position. These \lq bins\rq\ are located such that the shock lies 
at the border between two of them. 
An independent way of measuring the density is to count 
particles which cross a plane at a certain position. The count rate
is proportional to the flux through this plane; to obtain the density,  
one divides this quantity by the relative velocity of the binned 
particle and the plane. 
The shock plane and other planes at constant $\xi$ are stationary in the 
dHT frame, so that the particle velocity relative to the planes is 
 $v\mu\cos\phidht$ in the upstream region (see footnote \ref{dht_trans} for 
the relation between $\Phi$ and $\phidht$), where the particle velocity $v$ 
is multiplied by 
the cosine of the angle with respect to the magnetic field $\mu$ and the 
cosine of the inclination angle of the magnetic field relative to the normal
of the plane. The three filled dots at $\xi=0, 0.5$ and 1 show the density 
in the upstream region, whereas the filled square shows the density at the 
downstream side of the shock at $\xi=0$. In particular, the two values at
 $\xi=0$ are calculated from the integration of Eqs.~(\ref{n_int}) and 
(\ref{n_int_2}) over the dHT distributions shown in Fig.~\ref{pitch} 
(open circles). The normalisation is in each case such as to give unity 
far downstream of the shock.
\begin{figure}[t]
    \vspace{-3.1cm}
    \begin{center} 
      \epsfxsize9.0cm 
      \mbox{\epsffile{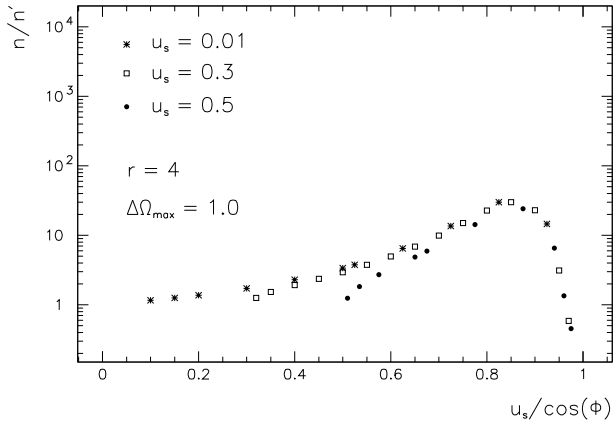}}
    \end{center}
    \vspace{-1.3cm} 
    \protect\caption{Ratio of the upstream to the downstream density at the 
          shock front vs. the intersection velocity of the shock and
          the magnetic field for {\em intermediate} pitch-angle scattering
          with $\dom =1.0$, compression ratio $r=4$ and three different 
          shock speeds. The values of $n/n'$
          are taken from flux measurements at the shock front.
          For values of $u_{\rm s}/\cos\Phi \protect\la 0.8$, the results are
          essentially the same as those shown in Fig.~\protect\ref{jump_us}.}
    \label{jump_us_do1}
\end{figure}

Both methods display a discontinuous density profile in 
Fig.~\ref{profile}. However, finite spatial resolution means that
the binning method systematically underestimates the value of the density 
when approaching the shock front from either side.
The second method, on the other hand, is a precise measure of the density
at $\xi=0$.
For the example of Fig.~\ref{profile} ($10^6$ independent particles)
we get $n/n'=3.641\pm 0.004$, where $n$ represents the filled circle, and $n'$ 
represents the filled square at $\xi=0$. 
Figure~\ref{jump_us} shows the ratio $n/n'$ for various shock velocities
and inclination angles. (The number of particles lies between $6\cdot 10^3$
to $5\cdot 10^6$ and the statistical error of $n/n'$ is $<5\%$). 
For $\Phi=0$ (parallel shock) we get a
continuous density ($n=n'$) for all shock speeds, because here no reflection
can occur (see Sect.~\ref{analytical}). 
If the velocity of the intersection point of the shock
and the magnetic field exceeds 0.8 of the particle velocity 
(in the case $v=c$ discussed here), the upstream density becomes more 
than 10 times the downstream density at the shock front. 
This ratio increases very rapidly for higher intersection velocities.
Our calculations are performed for test particles, but in reality such
particles may exert a substantial pressure, which 
would be important in calculations which include the back-reaction of
the particles on the flow, as pointed out by O91. 
\begin{figure}[t]
    \vspace{-1.0cm}
    \begin{center} 
      \epsfxsize9.5cm 
      \mbox{\epsffile{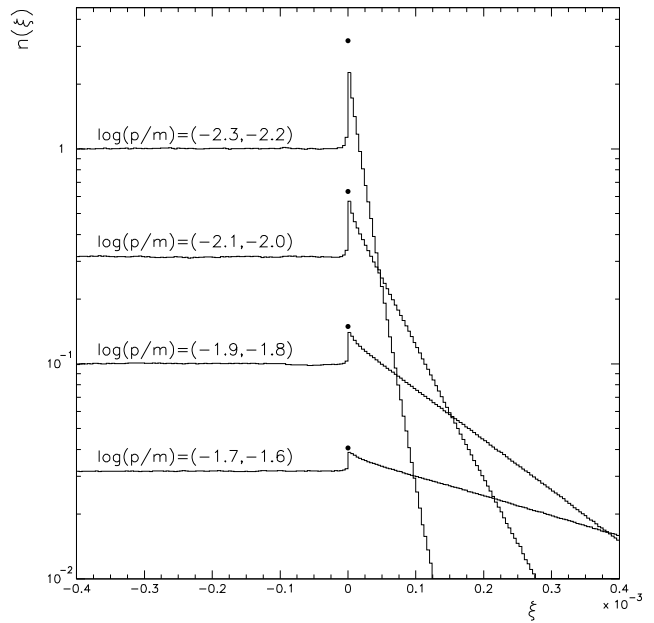}}
    \end{center}
    \vspace{-1.2cm} 
    \protect\caption{Monte-Carlo Simulation of the density profile at an 
            oblique shock ($\Phi=60^{\circ}$) with compression ratio $r=3.5$ 
            and shock speed $u_{\rm s}=300\,\mbox{km s}^{-1}$, for 
            various regions of the particle velocity. 
            The normalisation of each plot is chosen to avoid 
            overlap. The maximal scattering angle is $\dom=0.5$. 
            The filled dots show the upstream density measured 
            from the flux at the shock and normalised to the value 
            of the corresponding plot far downstream.}
    \label{dens_non_rel}
\end{figure}
However (as discussed in Sect.~\ref{spectral_index}),
increasing the magnitude of the maximal change in pitch angle ($\dom$) 
leads to a strong reduction of the highest possible upstream densities.
This can be seen very clearly from Fig.~\ref{jump_us_do1}, where
the maximal change in pitch angle is $\dom =1.0$,
and arises because in this case the angular distribution cannot
develop a sharply peaked structure such as that shown in
Fig.~\ref{pitch}. 
For isotropisation of the pitch angle after each scattering event, 
the ratio $n/n'$ for $r=4$ is always smaller than 2 for all inclination 
angles which are possible for subluminal oblique shocks ($u_{\rm s}<\cos\Phi$).

Finally, in Fig.~\ref{dens_non_rel} we show the density profile for an 
oblique shock ($\Phi=60^{\circ}$) with non-relativistic shock velocity
 $u_{\rm s}=300\,\mbox{km s}^{-1}$. We have simulated the acceleration of test
particles at an unmodified low Mach-number shock with compression 
ratio $r=3.5$, typical for shocks in the solar wind.
The pitch-angle scattering is described by 
Eqs.~(\ref{scatt_1}) and (\ref{scatt_2})
with $\dom=0.5$. This plot shows the density for 4 different velocity
regimes of the accelerated particles, which were injected with velocity
 $p/m= 2.1\cdot 10^{-3}$, just higher than the intersection velocity of 
 shock and magnetic field ($u_{\rm s}/\cos\Phi=2\cdot 10^{-3}$).
The density is normalised such that overlap of the four plots is avoided. 
For each plot in Fig.~\ref{dens_non_rel}
the filled dot represents the density measured at the shock front on the 
{\em up}stream side from the flux at the shock plane 
(compare Fig.~\ref{profile} and the corresponding text). For the lowest 
energy particles shown in Fig.~\ref{dens_non_rel}, which would correspond
to protons with a kinetic energy between 12 and 19~keV, the density 
upstream of the shock exceeds  the density far downstream by more than a 
factor of 3, and the ratio of the density upstream relative to downstream
 {\em at} the shock front is given by $n/n'= 2.81\pm 0.02$ at this
 \lq energy band\rq\ (note that the error represents only the statistical 
fluctuation). It can be seen from Fig.~\ref{dens_non_rel} that these ratios 
depend on the velocity of the particles. If the particle velocity exceeds the 
velocity of shock and magnetic field intersection by an order of magnitude, 
the distribution becomes more isotropic and therefore the density jump 
tends to disappear, as discussed in Sect.~\ref{contro} and \ref{analytical}.
\section{Discussion}
\label{disc}
We have presented an analysis of particle acceleration at oblique shocks,
with special emphasis on the density of accelerated test particles. 
It was shown analytically that in general a density jump can occur at an
oblique shock front. This was done on the basis of Liouville's theorem and
by integrating the phase space density, without the a priori assumption of
an isotropic pitch-angle distribution. It turns out that for situations in
which the pitch-angle distribution is non-isotropic, a density peak at the
shock front can appear. This peak has a discontinuity at the position of
the shock, if the adiabatic treatment is used. 
One can describe the resulting precursor
of upstream accelerated particles as due to reflections at the shock front.

We used Monte-Carlo simulations to calculate the
spectral index and the pitch-angle distribution. These results are in 
good agreement with semi-analytical calculations from K\&H89.
The corresponding density profile shows a pronounced
discontinuity for a large range of parameters, which was also 
found by O91 (see Fig. 6 therein), and our results are quantitatively
in very good agreement ($r=5.28$, $u_{\rm s}=0.3$, $\dom =0.1$,
 $u_{\rm s}/\cos\Phi=0.6 \Rightarrow n/n'= 6.31\pm 0.04$; and 
$u_{\rm s}/\cos\Phi=0.8771 \Rightarrow n/n'= 393\pm 7$).
This discontinuity results from the larger density of reflected 
particles as compared to 
transmitted ones than is found in the isotropic case.
Especially in the case of infinitesimal pitch-angle
scattering, the particles undergo repeated reflections (by which they are 
accelerated) before they reach a pitch angle at which they are 
able to cross the shock into the downstream region.
 
Allowing for a larger maximum value in the change of the pitch angle increases
the probability of entering the loss cone per scattering event, and therefore 
crossing the shock from upstream to downstream.
This reduces the density contrast (Fig.~\ref{jump_us_do1}).
The effect on the spectral index is restricted to a small region in the
parameter $u_{\rm s}/\cos\Phi$ (the intersection velocity of magnetic field and
shock) (Fig.~\ref{index_do1}). The reason for this steepening is that the 
acceleration due to reflections becomes less effective. For large 
pitch-angle scattering where the pitch angle is randomised after every 
scattering event, the pile-up effect is almost absent, and the minimal 
spectral index (the flattest spectrum) which can be reached at oblique 
shocks with non-relativistic shock velocities is $s\ga 3.4$ (see also N\&T95).  

In the case of non-relativistic particles accelerated at solar system
shocks, a significant density peak can occur only for highly oblique magnetic 
fields for particles whose velocity is less than an order of magnitude greater
than the intersection velocity of shock and magnetic field.
The occurrence of a density peak of accelerated particles at the 
shock front could in principle be detected in situ by space observations in 
the solar system, so that it is important to determine whether effects such 
as modification of the velocity profile by the 
pressure of accelerated particles can affect our results. 
Simulations of this nonlinear problem have been performed by 
Ellison et al.~(\cite{ElBaJo96}). However, they detect no difference in the 
spectral index 
between runs with infinitesimal and large pitch-angle scattering, whereas we 
predict that a difference should 
accompany a high density peak ahead of the shock front, as discussed in 
Sec.~\ref{spectral_index}. 
Very recently, numerical solutions of the transport equation for 
mildly relativistic particles at solar system shocks have found
a density peak at the shock front (Ruffolo~\cite{Ruff99}).

A pile-up of electrons 
ahead of the blast-wave associated with a supernova remnant could cause a 
synchrotron flux which {\em decreases} downstream of the shock instead of 
increasing due to the compression of the magnetic field (O91). 
Whereas this could always happen at a shock which propagates in an 
inhomogeneous medium, such a signature could be also produced by a shock 
moving in a homogeneous interstellar medium with an oblique 
magnetic field as a result of a pile-up of reflected particles.

The synchrotron emission as a function of frequency scales as
 $\epsilon(\nu) \propto n \,\nu^{-\alpha}\,B^{\alpha+1}$, 
where $\alpha=(s-3)/2$. 
Because of the enhanced synchrotron emission due to the increasing 
magnetic field downstream of the shock, the observation of the
density peak upstream of an oblique shock would be possible only 
for very high obliquities observed at high angular resolution. 
For a typical supernova shock velocity of 
 $u_{\rm s}=7000\,\mbox{km s}^{-1}$, an upstream inclination angle of
 $\Phi\ga 88^{\circ}$
 leads to an upstream density (of highly relativistic particles) which 
exceeds by more than a factor 10 the downstream density $n/n'\ga 10$ 
(see Fig.~\ref{jump_us} for infinitesimal, and Fig.~\ref{jump_us_do1} for 
intermediate pitch-angle scattering). At the same time the very flat spectrum
produced by these particles ($s\approx 3$) leads to only a linear 
enhancement of the downstream synchrotron emission due to the compressed 
magnetic field, $(B/B')^{\alpha+1}\approx 1/4$ (for $r=4$). This means that 
the synchrotron emission upstream would exceed the emission downstream.
We estimate that such an effect could
be resolved in the radio range and for parameters typical of a young SNR
such as Tycho if the scattering frequency
$\nu_{\rm s}$, defined in Eq.~(\ref{kappa}), is smaller than
the gyro-frequency by a factor of roughly $10^3$.

Observations of the Tycho supernova remnant have revealed a density peak
in the vicinity of the blast wave (Reynoso et al.~\cite{Reetal97}), 
which cannot be understood by the theory of an unmodified parallel shock 
moving in a homogeneous medium. At least {\em one} of these assumptions 
must be dropped in a model of that part of the remnant which shows this 
feature. A highly oblique shock is a possibility, especially if nonlinear 
effects in the precursor cause preferential 
alignment of the field in the plane of the shock, as suggested by recent
simulations (Bell, private comm.).
 However, the observed spectral indices of young shell-like supernova 
remnants  are in the range 
 $0.5\la\alpha\la 0.8$ (Reynolds \& Ellison \cite{ReEl92}), which corresponds
to $4\la s\la 4.6$ and deviates from the standard result for compression
 $r=4$ at a highly oblique shock. Spectral indices in this regime
can be produced by stochastic perpendicular magnetic fields
 (Kirk et al.~\cite{KiDuGa96}; Gieseler~\cite{Doc98}) or strongly modified
oblique shocks (Reynolds \& Ellison \cite{ReEl92}). Whereas in the former
case a density peak ahead of the shock does not arise (Gieseler~\cite{Doc98}), 
in the latter 
case the pile-up of accelerated particles may be an important effect.
Alternatively, nonlinear hydrodynamic effects of the accelerated particles 
can create an unstable density spike 
downstream of the shock, as shown by Jun \& Jones~(\cite{JuJo97}), which may 
be responsible for the enhancement in the synchrotron emission.

We have shown that highly oblique shocks can produce
a pronounced density peak due to a pile-up of accelerated 
particles ahead of the shock front. In Monte-Carlo simulations, 
this might appear as a discontinuity or even be overlooked, depending on the 
method used and the spatial 
resolution achieved. We point out that the nonlinear effects of such a 
pile-up could be significant, so that it is 
important to locate this effect in simulations which incorporate the reaction 
of the pressure of accelerated particles on the plasma dynamics. 

\begin{acknowledgements}
This work was supported by the European Commission under the TMR programme,
contract number FMRX-CT98-0168. U.D.J.G. acknowledges support from the 
Deutsche Forschungsgemeinschaft under SFB~328.
\end{acknowledgements}

\end{document}